\documentclass[twocolumn,showpacs,preprintnumbers,prb,amsmath,amssymb]{revtex4}
\usepackage{dcolumn}
\usepackage{bm}
\usepackage{amsmath,amssymb}
\usepackage{epsfig}
\usepackage{subfigure}
\usepackage{epsf}
\usepackage{graphicx}
\begin{document}
\title{Memory and aging effects in antiferromagnetic nanoparticles}
\author{Sunil Kumar Mishra}
\address{Department of Physics, Indian Institute of Technology Kanpur-208016, India}
%\date{\today}
%\ead{sunilkm@iitk.ac.in}
\begin{abstract}
We investigate slow dynamics of collection of a few antiferromagnetic NiO nanoparticles. The zero-field cooled magnetization exhibits  size dependent fluctuations.  We find memory effects in field cooled magnetization, as well as aging effects in thermoremenant magnetization of antiferromagnetic nanoparticles. The antiferromagnetic nanoparticles show a stronger memory effect than the corresponding effect in the ferromagnetic particles, when the distribution of particles include very small sizes. The situation reverses for bigger sizes. The relaxation of the magnetization after a sudden cooling, heating and removal of fields reiterate the memory effects. We also see a weak signature of size-dependent magnetization fluctuations in aging effect of antiferromagnetic nanoparticles. We find a two-step relaxation of thermoremenant magnetization in antiferromagnetic case, which differs qualitatively from relaxation of ferromagnetic nanoparticles. 
%We also study the role of size distribution on the memory and aging effects by considering single peak size and two peak size distribution.
\end{abstract}

\maketitle
%%%%%%%%%%%%%%%%%%%%%%%%%%%%%%%%%%%%%%%%%%%%%%%%%%%%%%%%%%%%%%%%%%%%%%%%%%%
\section{Introduction}
%%%%%%%%%%%%%%%%%%%%%%%%%%%%%%%%%%%%%%%%%%%%%%%%%%%%%%%%%%%%%%%%%%%%%%%%%%%
 The interest in the study of magnetism in nanoparticles has been renewed from the last few decades due to their technological \cite{weller,richter,berry} as well as fundamental research aspects. \cite{fiorani,kodama,kodama1,jonsson1,jonsson2,sahoo1,dormann,labarata,jonsson3,djuberg,malay2, garcia1,jonsson,mamiya,sahoo2,sun,sahoo3,zheng,sasaki,tsoi,malay,wang,du,suzuki,winkler,makhlouf,tiwari,sunil}
% \cite{fiorani,kodama,kodama1,jonsson1,sahoo1,dormann,jonsson2,jonsson3,labarata,djuberg,jonsson,sahoo2,sun,sahoo3,zheng,sasaki,tsoi,malay,wang,du,suzuki,mamiya,winkler,makhlouf,tiwari,sunil}.
The magnetic properties of nanoparticles are dominated by finite-size effects, and the surface anomalies such as
surface anisotropy and roughness. \cite{fiorani,kodama,kodama1} As the particle size decreases, the fraction of
the spins lying on the surface of a nanoparticle increases, thus, making the surface play an important role. The
reduced coordination of the surface spins causes a symmetry lowering locally, and leads to a surface anisotropy,
that starts dominating as the particle size decreases.

The dynamics of an assembly of nanoparticles at low temperatures gained a lot of attention over the last few years. In a dilute system of nanoparticles, the interparticle interaction is very small as compared to the anisotropy energy of the individual particle. These isolated particles follow the dynamics in accordance with N\'eel-Brown model \cite{brown} and the system is known as superparamagnetic. The giant spin moment of nanoparticles
thermally fluctuates between their easy directions at high temperatures. As the temperature is lowered towards a blocking temperature, the relaxation time becomes equal to the measuring time and the super spin moments freeze along one of their easy directions. As the role of interparticle interaction becomes significant, nanoparticles do not behave like individual particles, rather their dynamics is governed by the collective behavior of the particles, like in a spin glass. \cite{jonsson1,jonsson2,sahoo1,dormann,labarata,jonsson3} 
   This super spin-glass phase has been characterized by observations of a critical 
slowing down, \cite{djuberg,malay2,garcia1} a divergence in
the nonlinear susceptibility, \cite{malay2,garcia1,jonsson} and aging and relaxation effects in the low-frequency ac susceptibility. \cite{mamiya}
 The Monte Carlo simulations on the system of assembly of nanoparticles show aging \cite{andersson} and magnetic relaxation behavior \cite{ulrich} like in a spin glass, but simulations of zero field cooling (ZFC) and field cooling (FC) susceptibilities show no indication of spin-glass ordering. \cite{garcia}
The aging
 and memory effects are the two aspects that have been studied extensively in recent years, however mostly for the case of ferromagnetic particles. \cite{jonsson1,jonsson2,sahoo1,jonsson,sahoo2,sun,sahoo3,zheng,sasaki,tsoi,malay,wang,du,suzuki} 
In this paper our aim is to investigate these effects for collection of a few antiferromagnetic nanoparticles (AFNs) which has received relatively lesser attention. 

 N\'eel predicted \cite{neel,jacob} that AFNs exhibit weak ferromagnetism 
and superparamagnetism behavior, which is attributed to a net magnetic moment due 
to incomplete magnetic compensation between the atoms in two sublattices. These sublattices are identical in every
respect, except that the atomic moments in one sublattice are antiparallel to that in other.
For Nickel Oxide (NiO) nanoparticles, the magnetic moment $\mu$ predicted by N\'eel's model varies as $ \mu \sim n^{\frac{1}{3}} \mu_{Ni^{2+}} $, where $n$ is the number of spins. \cite{richard2} 
Weak ferromagnetism and superparamagnetism were later confirmed by experiments \cite {schuele} on fine particles and extremely
fine particles of NiO, respectively. Mishra and Subrahmanayam \cite{sunil} have also concluded a net magnetic moment in NiO AFN due to finite size effect in N\'eel-state ordering, which is showing a non-monotonic and oscillatory dependence on particle size $R$. The amplitude of fluctuations were found to be varying linearly with $R$, consistent with N\'eel's model. However, the value of net magnetic moment due to finite size effect in N\'eel-state ordering
 does not quantify the large magnetic moment experimentally observed in NiO nanoparticles. This discrepancy is due to the ignorance of different ordering of surface spins than the core spins in these studies. 
Monte Carlo simulations on antiferromagnetic particles highlighted the dominant role of surface spins in net magnetization of the nanoparticles. \cite{trohidou,zianni}
The surface effects have been considered to be the major cause for the large magnetic moment
in the NiO nanoparticle. \cite{fiorani, kodama1, winkler,sunil,zysler} The breakdown of the dominant next-nearest neighbor antiferromagnetic interaction on the
surface of the nanoparticle leads to uncompensated 
spins. These uncompensated spins play a vital role in determining the magnetic behavior of NiO 
nanoparticles.
Thus, an enhancement of surface and interface effects make the AFNs an interesting 
area of research. \cite{fiorani,kodama,kodama1,winkler,makhlouf,tiwari}

 Recently, Mishra and Subrahmanyam \cite{sunil} have shown that for NiO nanoparticles the net magnetic moment, a combined effect of
 surface roughness effect and finite-size effects in core magnetization, exhibits size dependent fluctuations in net magnetic moment. 
These size dependent fluctuations in magnetization lead to a dynamics which is qualitatively different from ferromagnetic nanoparticles.

 In this paper, we mainly focus
on the dynamics of collection of a few noninteracting NiO nanoparticles using master equation approach. %We will compare the Monte Carlo time step with real time step by comparing the Monte Carlo study with analytical calculation. 
We examine the effect of the size-dependent magnetization fluctuations on the time-dependent properties of sparse assembly of the nanoparticles. We will compare the ZFC and FC magnetizations of antiferromagnetic particles with the ferromagnetic case. We will consider the  effect of polydispersity as well. We will perform a series of heating/cooling processes which were earlier discussed in the case of ferromagnetic nanoparticles. \cite{sun,sasaki,tsoi,malay} We will also discuss the memory effects and aging effects throughly. The organization of this paper is as follows. 
In section \ref{mod1} we discuss the model. The ZFC and FC magnetizations for various distributions are discussed in section \ref{seczfc} and in section \ref{secmemory} we show the memory effects investigations. Aging effects has been presented in section \ref{secaging}. Finally we summarize in Sec \ref{secconclusion}.

%%%%%%%%%%%%%%%%%%%%%%%%%%%%%%%%%%%%%%%%%%%%%%%%%%%%%%%%%%%%%%%%%%%%%%%
\section{Relaxation in superparamagnets and polydispersity}
\label{mod1}
%%%%%%%%%%%%%%%%%%%%%%%%%%%%%%%%%%%%%%%%%%%%%%%%%%%%%%%%%%%%%%%%%%%%%%%%%
\begin{figure}[t]
%\vspace*{1.2cm}
\begin{center}
\includegraphics [angle=0,width=8cm] {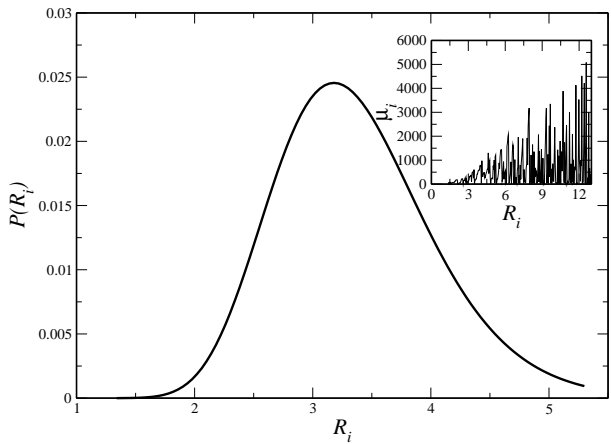}
\caption{ A log normal distribution of nanoparticles of sizes ranging $1.3a_0-5.3a_0$. The inset displays the magnetic moment vs particle size. Magnetic moment has a non-monotonic and oscillatory dependence on $R$. }
\label{fig1a}
\end{center}
\end{figure}
  In an earlier study, \cite{sunil} we have shown that the total magnetic moment of NiO antiferromagnetic nanoparticle displays size dependent fluctuations as shown in the inset of figure \ref{fig1a}. The net magnetic moment shows a trend where magnetic moment is very small for smallest size particles. Increasing the size, magnetic moment increases and reaches a maximum at $R\sim 10 a_0$, and again decreases towards the bulk value. 
On the other hand, the net magnetic moment  of ferromagnetic nanoparticles \cite{sasaki}  shows a linear dependence on the size of the particles
%In the present study for the AFN, the magnetic moment does not depend linearly on size, instead it shows size dependent fluctuations.
Hence one might expect the role of these size-dependent fluctuations in magnetization should be manifested in the time dependent properties of antiferromagnetic nanoparticles.

In our simple model, the energy of each particle $i$
is contributed by anisotropy energy (either due
to the shape or the crystalline structure of the particle),
and Zeeman energy. For the sake of simplicity we assume that the direction of field is same as that of anisotropy axes. Thus
\begin{eqnarray}
E_i=-\mathcal{K}V_i+\mathcal{H} \mu_i,
\end{eqnarray}
where $\mathcal{K}$ is the anisotropy constant and $\mathcal{H}$ is the applied magnetic field.
In the absence of field, the
superparamagnetic relaxation time for the thermal activation over the energy barrier $\mathcal{K}V_i$ is given by $ \tau =\tau_0 \rm{exp} (\mathcal{K}V_i/k_{\rm{B}} T)$,
  where $\tau_0$, the microscopic time is of the order of $10^{-9}$ sec and $k_{\rm{B}}$ is Boltzmann constant. The anisotropy 
constant $\mathcal{K}$ has a typical value \cite{hutchings} about $4\times10^{-1}$ J${\rm{cm}}^{-3}$.
  The occupation probabilities with the magnetic moment parallel and antiparallel
to the magnetic field direction are denoted by $p_1(t)$ and $p_2 (t) = 1 - p_1(t)$ ,
respectively. These probabilities must satisfy the
master equation\cite{sasaki}
\begin{eqnarray}
  \frac{d}{dt}p_1(t) = - \lambda_{12}(t) p_1(t) + \lambda_{21}(t) (1 - p_1(t)),
\label{mseqn}
\end{eqnarray}

where the parameters $\lambda_{12}(t)$ and $\lambda_{21}(t)$ are the rate of transition of the magnetic moment from the two states at time $t$, given as ${\tau_0}^{-1} \rm{exp} \left[ -\mathcal{K}V_i/T(t)\right] \left[  1-{\mu}_i h(t)/T(t)\right]  $ and ${\tau_0}^{-1} \rm{exp} \left[ -\mathcal{K}V_i/T(t)\right] \left[  1+{\mu}_i h(t)/T(t)\right] $ respectively. 
Using the values of parameters $\lambda_{12}(t)$ and $\lambda_{21}(t)$, Eq. \ref{mseqn} can be simplified as
\begin{eqnarray}
  \frac{d}{dt}p_1(t)=-\frac{1}{\tau(t)}p_1(t)+\frac{1}{2\tau(t)}\left[  1+\frac{{\mu}_i h(t)}{T(t)}\right]. 
\label{mseqn1}
\end{eqnarray}
%   The above master equation
% can be solved analytically for any 
%temperatures and field protocols represented by $T(t)$ and $h(t)$
%from a given initial condition, 
The magnetic moment of the
particle of volume $V_i$ given by
\begin{eqnarray}
                {\mu} (t,V_i) = \left[ 2p_1( t,V_i) - 1\right]  \mu_i.
\label{mtv}
\end{eqnarray}

For $\mathcal{H}(t) = \mathcal{H} $ and $T(t) = T$, we can write
\begin{eqnarray}
 {\mu} (t,V_i)=\mu (0,V_i) \rm{exp} (-t/\tau)+\frac{{\mu_i}^2 \mathcal{H}}{T}\left\lbrace 1-\rm{exp}(-t/\tau) \right\rbrace. 
\label{fixedt}
\end{eqnarray}
For a constant magnetic field, the above equation governs the relaxation of magnetization at each temperature step.
Thus the total magnetic moment of the system of nanoparticles with volume distribution $P(V_i)$ is given by
\begin{eqnarray}
 \mu(t)=\int {\mu} (t,V_i) P(V_i) d V_i.
\label{mtvpv}
\end{eqnarray}
\begin{figure*}
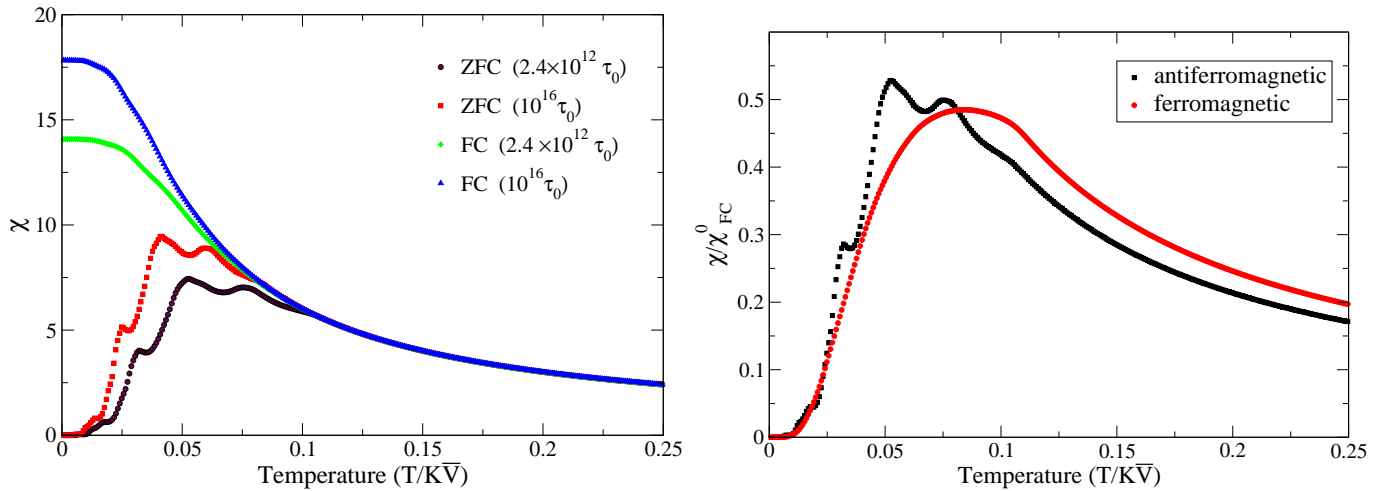

\hspace*{-.8cm}
\centering
\begin{tabular}{cc}
\epsfig{file=fczfc.eps,width=0.5\linewidth,clip=} &
\epsfig{file=ferro_cmp.eps,width=0.5\linewidth,clip=} 
\end{tabular}
\caption{a)ZFC and FC susceptibilities have been plotted from the solution of Eq. \ref{mtvpv} with a heating/cooling rate $10^{12}\rm{\tau}_0$ and $10^{16}\tau_0$ per temperature step. Ripples can be seen in ZFC susceptibilities. Increasing heating rate lowers the blocking temperature. b) ZFC susceptibilities for ferromagnetic and antiferromagnetic cases using same size distribution with heating rate $10^{12} \tau_0$ has been shown. The susceptibilities are normalized with $\chi_{\rm{FC}}$ at $T=0$. It can be seen that ripples are absent in the ferromagnetic case.}
\label{analy}
\end{figure*}
 The size-distribution plays a significant role in the overall dynamics of the system of nanoparticles  governed by Eq. \ref{mtvpv}. As we can see that due to the exponential dependence of $\tau$ on the particle size $V_i$, even a weak polydispersity may lead to a broad distribution of relaxation times, which gives rise to an interesting slow dynamics. 
For a dc measurement, if relaxation time coincides with the measurement time scale $\tau_m$, we can define \cite{bean} a critical volume $V_{\rm{B}}$ as $\mathcal{K}V_{\rm{B}}=k_{\rm{B}} T_{\rm{B}} \rm{ln}(\tau_m/\tau_0)$, where $T_{\rm{B}}$ is referred as blocking temperature.
The critical volume $V_{\rm{B}}$ has strong linear dependence on $T_B$ and weakly logarithmic dependence on the observation time scale $\tau_m$. If the volume of the particle $V_i$ in a 
polydisperse system is less than $V_{\rm{B}}$, the super spin would have undergone many rotations within the measurement time scale with an average magnetic moment zero. These particles are termed as {\it superparamagnetic} particles. On the other hand if $V_i > V_{\rm{B}}$, the super spins can not completely rotate within the measurement time window and show {\it blocked or frozen} behavior. However, the particles having volume $V_i\simeq V_{\rm{B}}$ are in {\it dynamically active} regime.
%Using equation (\ref{mseqn}) and (\ref{mtv})
%In our numerical study,
%we have performed Monte Carlo simulations using $10^4$ particles.
The systems of magnetic nanoparticles are in general polydisperse. The shape and size of the particles are not well known but the particle size distribution is often found to be lognormal. \cite{buhrman}
We consider the system consisting of lognormally distributed, widely dispersed nanoparticles, hence non interacting among each other. 
 The volume $V_i$ of each particle is obtained from a log normal distribution 
\begin{eqnarray}
P(V_i;\sigma;\upsilon) = \frac{1}{\sigma V_i \sqrt{2\pi}} \rm{exp} \left[  \frac{-(ln(V_i)-\upsilon)^2}{2\sigma^2}\right],
%P(V_i)	= \frac{1}{\sigma V_i \sqrt{2\pi}} e^\frac{-(ln(V_i)-\mu)^2}{(2\sigma^2)}.
\end{eqnarray}
where $\upsilon=ln(\bar{V})$, $\bar{V}$ is the mean size and $\sigma$ the width of the distribution. The distribution consists of $10^4$ particles of sizes between $R =1.3a_0$ and $R=5.3a_0$, where $a_0(=4.17\AA)$ is the lattice parameter of NiO. \cite{hutchings} The total number of particles are purposely chosen to be small in order to see the effect of size dependent magnetization fluctuation on the relaxation dynamics of assembly of nanoparticles. 
 We can solve Eqs. \ref {mseqn1}, \ref{mtv} and \ref{mtvpv} for any heating/cooling process. For example, we can numerically solve these equations for a zero field cooled (ZFC) protocol. In a genuine ZFC protocol, system is cooled from a very high temperature to lowest temperature in the absence of magnetic field. Thus system is demagnetized at the lowest temperature. This condition is analogous to  $p_1(0)=1/2$ in Eq. \ref{mtv}. Now a constant field is applied and the system is heated upto high temperature. At each temperature change we evolve the system using Eq. \ref{fixedt}. We can define heating rate in the process as total time elapsed at each temperature change. Thus heating/cooling rate $10^{12}\tau_0$ per temperatue unit corresponds to heating/cooling process in which system is relaxed for $t=10^{12}\tau_0$ at each temperature step.

In this paper, volume $V_i$ is measured in the units of average volume $\bar{V}$. Also the average anisotropic energy $\mathcal{K}\bar{V}$ is taken as the unit of energy, which is equal to $480\rm{K}$ for $\bar{V}=193 {a_0}^3$. Thus by setting $k_{\rm{B}}=1$, we use $\mathcal{K}\bar{V}$ as a unit of temperature $T$ and and field $\mathcal{H}$. Hereinafter we use a dimensionless quantity $h={\mu}_B \mathcal{H}/\mathcal{K} a_0^3$ as a unit of field, e.g., $h=0.01$ is equivalent to a magnetic field $300$ Gauss.
%%%%%%%%%%%%%%%%%%%%%%%%%%%%%%%%%%%%%%%%%%%%%%%%%%%%%%%%%%%%%%%%%%%%
\section{ZFC and FC magnetizations }
\label{seczfc} 
%%%%%%%%%%%%%%%%%%%%%%%%%%%%%%%%%%%%%%%%%%%%%%%%%%%%%%%%%%%%%%%%%%%%
The relaxation phenomenon of nanoparticles is often investigated using two different protocols, ZFC magnetization and FC magnetization measurements. %The time dependent magnetization can be obtained by solving the master equation \cite{sasaki}
%\begin{eqnarray*}
In a ZFC magnetization measurement, the system is first demagnetized at a very high temperature and then cooled down to a low temperature in a zero magnetic field. A small
magnetic field is then applied and  the magnetization
is calculated as a function of increasing temperature.
 We have shown a plot of FC-ZFC magnetization with temperature for polydisperse NiO 
nanoparticles in Fig. \ref{analy}(a) using Eqs. \ref {mseqn1}, \ref{mtv} and \ref{mtvpv} with heating/cooling rates $2.4\times10^{12} \tau_0$ and $10^{16}\tau_0$ per temperature step. For a ZFC process we find that increasing the temperature susceptibility $\chi_{\rm{ZFC}}$ increases, attains a maximum value  
at a blocking temperature ($T_{\rm{B}}$), and then starts decreasing.  We see that the average blocking temperature depends on the heating rate. Increasing the heating rate lowers the blocking temperature. For example, for $2.4\times10^{12} \tau_0$ per temperature unit heating rate, blocking temperature is $0.053$, whereas for  heating rate $10^{16} \tau_0$ per temperature unit, blocking temperature is $0.041$.
\begin{figure}[t]
%\vspace*{1.2cm}
\begin{center}
\includegraphics [angle=0,width=8cm] {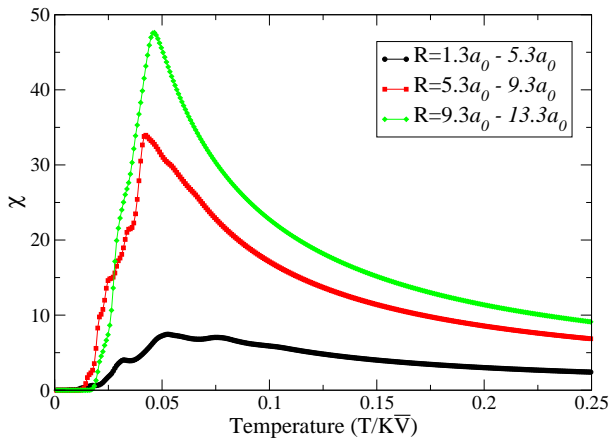}
\caption{ ZFC susceptibility for various size distributions is shown. The role of size dependent magnetization fluctuations can be explicitly seen as a ripple in the curve. For bigger sizes distribution $R=9.3a_0-13.3a_0$, the curve is smoother.}
\label{distcmp}
\end{center}
\end{figure}

In Fig. \ref{analy}(b), we have shown the comparison of ZFC magnetization for antiferromagnetic particles with the ferromagnetic particles of same sizes, where susceptibility has been normalized by FC value at $T=0$. The overall ZFC magnetization behavior is same for both cases except the presence of ripples in ZFC magnetization for antiferromagnetic particles. These ripples are attributed due to the size dependent fluctuations in magnetization. As we increase the heating rate these ripples become more pronounced. This can be seen in Fig. \ref{analy}{a}, where increasing the heating rate enhances the ripple in magnetization. We have also investigated the effect of polidispersity by incorporating various size distributions. We consider three distributions {\it dist.A}, {\it dist.B} and {\it dist.C} in which sizes are in the range of $1.3a_0-5.3a_0$, $5.3a_0-9.3a_0$ and $9.3a_0-13.3a_0$ respectively. The mean sizes in these distributions are $3.5a_0$, $7.5a_0$ and $11.5a_0$ respectively. We can argue from Fig. \ref{distcmp}, that the smaller size distributions exhibit more effects of fluctuations in magnetizations than the bigger sizes. For bigger sizes, we expect a little role of magnetization fluctuation as comapared to smaller sizes, which indeed is the case for the size range ($9a_0-13a_0$) as shown in Fig. \ref{distcmp}. We find that for the size range ($9a_0-13a_0$), ZFC curve is almost smooth, while for lower sizes, it exhibits ripples. 
                       
 During an FC measurement, the system is cooled
in the presence of a probing field from higher temperatures to a low temperature. 
%where temperature is changed in the steps of $\Delta T=.008 K \bar{V}/k_{\rm{B}}$ for every  $50$ 
%Monte Carlo steps. 
%In figure \ref{fczfc}(a), we have plotted FC and ZFC magnetic moment per particle using Monte Carlo method. 
We find that the FC susceptibility $\chi_{\rm{FC}}$ coincides with 
$\chi_{\rm{ZFC}}$ at higher temperature but departs from ZFC curve at lower temperatures, however well above the blocking temperature, and 
tends to a constant value with further lowering the temperature. The blocking temperature shows a substantial dependence on the heating 
rate. For infinitely slow heating rate, $T_{\rm{B}}$ approaches to zero and the ZFC curve shows similar behavior as FC curve. We also find that FC magnetization never decreases as the temperature is lowered which is a characteristic feature of superparamagnets. \cite{sasaki}
%%%%%%%%%%%%%%%%%%%%%%%%%%%%%%%%%%%%%%%%%%%%%%%%%%%%%%%%%%%%%%%%%%%%
\section{Memory effect}
\label{secmemory} 
%%%%%%%%%%%%%%%%%%%%%%%%%%%%%%%%%%%%%%%%%%%%%%%%%%%%%%%%%%%%%%%%%%%%
    Recently Sun {\it et al.} \cite{sun} have reported a striking result showing memory effects in the dc
magnetization
by a series of measurements on a permalloy $\rm{Ni_{81}Fe_{19}}$ nanoparticle sample. 
These measurements include FC and ZFC relaxations under the influence of temperature and magnetic field.  They cooled the sample in $50$ \rm{Oe} field at a constant cooling rate of $2$ \rm{K} per minute from $200$ \rm{K} to $T_{\rm{base}} =10$ \rm{K}. After reaching at $T_{\rm{base}}$, the sample was 
heated continuously at the same rate upto $200$ $\rm{K}$. The $M(T)$ curves thus obtained are
the normal FC curve which is referred by them as reference curve. The sample is cooled again at the same rate, but
with temporary stops at $T = 70$ $\rm{K}$, $50$ \rm{K} and $30$ \rm{K}
below blocking temperature $T_{\rm{B}}$ with a wait time $t_{\rm{w}} = 4$ hours at each stop. During each stop, the field was also turned off to let the magnetization relax.
After each pause the magnetic field was reapplied
and cooling was resumed. The cooling procedure produces a
steplike $M(T)$ curve. After reaching $T_{\rm{base}}$, the
sample is warmed continuously at the same rate to $T_{\rm{H}}$ in the
presence of the $50$ \rm{Oe} field. The $M(T)$
curve thus obtained also shows the steplike behavior around each stops.
Sun {\it et al} suggested that this `memory effect' indicates the possibility of hierarchical organization of metastable states resulting from interparticle interactions. Since hierarchical organization requires a large number of degree of freedom to be coupled, the memory effect may not arise due to the thermal relaxation of independent particle. 
\begin{figure*}
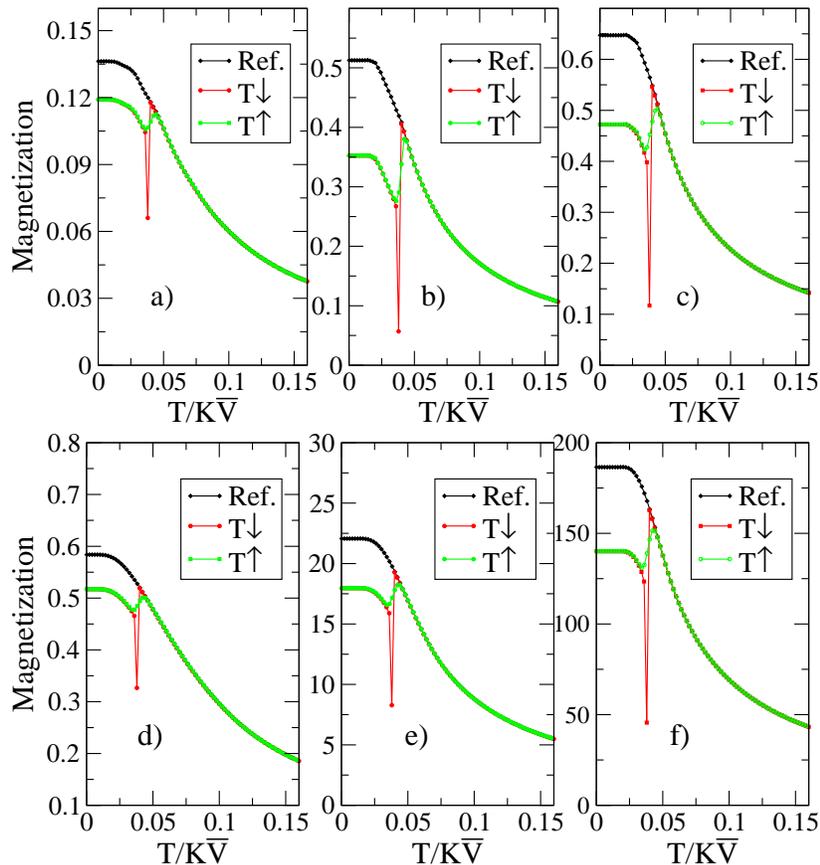

\hspace*{-.8cm}
\centering
\begin{tabular}{cc}
\epsfig{file=af_mry_cmp.eps,width=0.6\linewidth,clip=} \\
\epsfig{file=f_mry_cmp.eps,width=0.6\linewidth,clip=} 
\end{tabular}
\caption{Memory effect in the dc magnetization with a stop at $T=0.039$. Memory effects shown in (a), (b) and (c) correspond to antiferromagnetic nanoparticles for {\it dist.A} ($1.3a_0-5.3a_0$), {\it dist.B} ($5.3a_0-9.3a_0$) and {\it dist.C} ($9.3a_0-13.3a_0$) with a stop at $T=0.039$, while (d), (e) and (f) correspond to ferromagnetic nanoparticles of respective size distributions.}
\label{mry}
\end{figure*}
However more recently, Sasaki {\it et al.} \cite{sasaki} and Tsoi {\it et al.} \cite{tsoi} have reported similar results for the noninteracting or weakly interacting superparamagnetic system of ferritin nanoparticles and $\rm{Fe_2O_3}$ nanoparticles respectively. In these studies, the dynamics of the system of nanoparticles are assumed to be governed by a broad distribution of particle relaxation times arising from the distribution of particle sizes and sample inhomogeneities.
Experiments on NiO nanoparticles by Bisht and Rajeev \cite{bisht} also confirms a weak memory effect in these particles.

Chakraverty {\it et al} \cite{malay1} have investigated the effect of polydispersity and interactions among the particles in an assembly of nickel ferrite nanoparticles embedded in a host non magnetic $\rm{SiO}_2$ matrix. They found that either tuning the interparticle interaction or tailoring the particle size distribution in nanosized magnetic system leads to important  application in memory devices. 

We perform a similar study in thermoremenant-magnetization (TRM) protocol as that by Sun {\it et al} \cite{sun}, as shown in Fig. \ref{mry}(a)-(h). Figures \ref{mry}(a), (b) and (c) correspond to system of antiferromagnetic nanoparticles with distributions {\it dist.A}, {\it dist.B} and {\it dist.C} and Figs \ref{mry}(d), (e) and (f) correspond to ferromagnetic nanoparticles of size distributions{ \it dist.A}, {\it dist.B} and {\it dist.C}.
%Here, Particle sizes in the size distributions {\it dist.A}, {\it dist.B} and {\it dist.C} are in the range of $1.3a_0-5.3a_0$, $5.3a_0-9.3a_0$ and $9.3a_0-13.3a_0$ respectively. The mean sizes in these distributions are $3.5a_0$, $7.5a_0$ and $11.5a_0$ respectively. 
We first cool the system from a very high temperature to $T_{\rm{base}}=0.0018$ with a field $h=0.01$ and then again heat to get the Ref. curves in Figs \ref{mry} (a)-(h) with a cooling/heating rate $2.4\times10^{12} \tau_0$ per temperature step. We again cool the system from a high temperature to $T_{\rm{base}}$ but with a stop of $10^{14}\tau_0$ at $T=0.039$. The field is cut during the stop. After the pause the field is again applied and the system is again cooled up to base temperature $T_{\rm{base}}$. The process is shown as $T\downarrow$ curves in Figs \ref{mry}(a)-(h). Finally we heat the system at the same rate as that of cooling without any stop, shown as $T\uparrow$ curves in Figs \ref{mry}(a)-(h). We find that the magnetization shows an upturn exactly around $T=0.039$. Further heating recovers the Ref. curve. 

\begin{figure*}
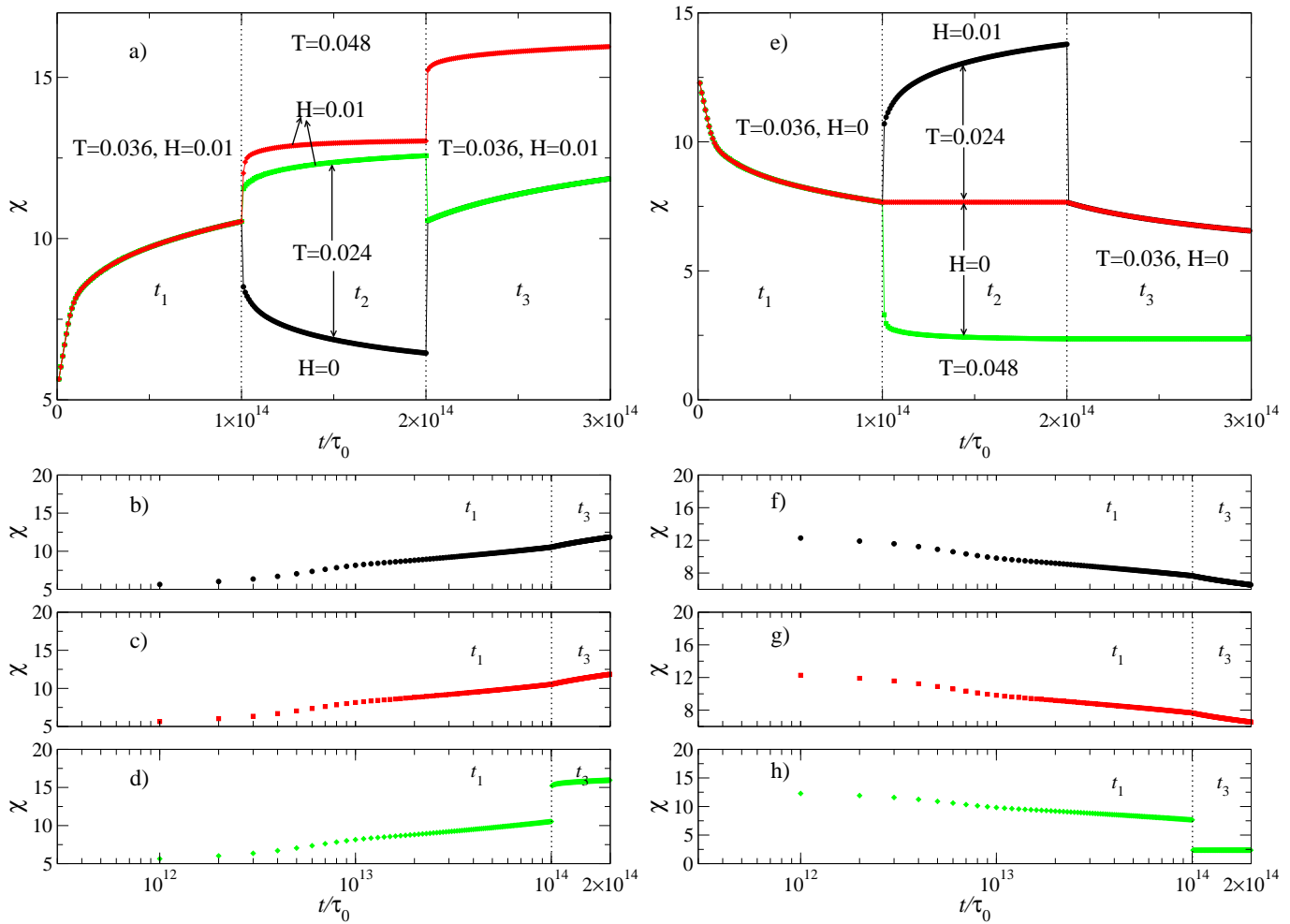

\hspace*{-.8cm}
\centering
\begin{tabular}{cc}
\epsfig{file=zfc_rlx.eps,width=0.50\linewidth,clip=} &
\epsfig{file=fc_rlx.eps,width=0.50\linewidth,clip=} \\
\epsfig{file=inset_zfc_rlx.eps,width=0.50\linewidth,clip=}& 
\epsfig{file=inset_fc_rlx.eps,width=0.50\linewidth,clip=}
\end{tabular}
\caption{(a) ZFC relaxation and (e) FC relaxation for a system of antiferromagnetic nanoparticles. Relaxation curves during $t_1$ and $t_3$ on lograthimic time scale for negative-temperature, field-change and positive-temperature cycles for ZFC case are shown in (b), (c) and (d) respectively and for FC case are plotted in (f), (g) and (h) respectively. }
\label{rlxzfc}
\end{figure*}

We can also shed some light on the role of polydispersity on memory effect. We compare the relative strength of memory effect in these distributions by introducing a parameter, memory fraction i.e., the ratio of $\Delta M /M^{\rm{Ref}}$ at the stop during memory measurement, where $\Delta M=M-M^{\rm{Ref}}$. In all the cases, we fix the waiting time at stop to be $10^{14}\tau_0$. The calculated values of the memory fraction for antiferromagnetic case is $13.3 \%$, $36.1\%$ and $27.39\%$ respectively. For ferromagnetic case the memory fraction takes values $11.4 \%$, $18\%$ and $25\%$ respectively. An interesting observation in the case of antiferromagnetic nanoparticle is the higher value of memory fraction for {\it dist.B} than that for {\it dist.A} and {\it dist.C}. As we know that the finite-size and surface roughness effects are responsible for the enhancement of net magnetization for the intermediate size ranges {\it dist.B}, the net memory dip is more in this case. However, for ferromagnetic nanoparticles, where the net magnetic moment enhances linearly with the size, we find an increasing trend of memory fraction. One more interesting aspect is that the memory fraction of antiferromagnetic nanoparticles are more than that of ferromagnetic nanoparticles in all the cases. Thus using the scale of memory fraction, we conclude that for smaller size distributions, memory effect is stronger in antiferromagnetic nanoparticles than ferromagnetic particles. However, it is obvious that for larger sizes distributions, memory effect in ferromagnetic particles exceeds to their antiferromagnetic counterpart. 

 We have examined the magnetization relaxation in NiO nanoparticles by a series of heating and cooling processes to understand the memory effects as discussed by sun {\it et al}. \cite{sun} 
 These relaxation studies are performed under both the ZFC and the TRM protocol. 

In a
ZFC protocol for the distribution {\it dist.A}, as shown in Fig. \ref{rlxzfc}(a), the sample is cooled down to $T_0=0.036$ in a zero field and a magnetic field $h=0.01$ is applied after zero wait time. The magnetization relaxation is calculated as a function of time for time interval $t_1=10^{14} \tau_0$ . After $t_1$ interval, we employ following three distinct routes, either changing temperature or applying magnetic field to study the relaxation for time $t_2 = 1\times10^{14}-4\times 10^{14} \tau_0$.

\begin{itemize}
%{\romannumeral 1}
\item[A]- Negative-temperature cycle: the sample is quenched in constant field to a lower temperature, $T_{\rm{low}}=0.024 $, and the magnetization is recorded for $t_2$ time interval.
\item[B]- Field-change cycle: the sample is quenched to a lower temperature, $T_{\rm{low}}=0.024$ with cutting off the magnetic field, and the magnetization is recorded for $t_2$ time period.
\item[C]- Positive-temperature cycle: the sample is quenched in constant field to a higher temperature, $T_{\rm{high}}=0.048$, and the magnetization is recorded again for $t_2$ time period.
\end{itemize}

At last we bring the system back to initial temperature $T_0=0.036$ and a constant field $h=0.01$ and find the relaxation for interval $t_3=2\times10^{14}-3\times 10^{14} \tau_0$.
% We have shown a relaxation curve with the
%ZFC method at $T_0=0.036$ for distribution {\it dist.A} in Fig. \ref{rlxzfc}(a). 
We find that during negative-temperature cycle for interval $t_2$, the particles which were dynamically active at $T_0$ becomes frozen at lower temperature $T_{\rm{low}}=0.036$ and the smaller particles which should be dynamically active at $T_{\rm{low}}$ have already been polarized during interval $t_1$, hence the relaxation becomes very weak during $t_2$ and a flat relaxation can be be observed in the figure. 
% for two peak polydisperse system in FIG \ref{rlxzfc}(b).
When the temperature returns to $T_0$, the
magnetization also comes back to the level it reached before
the temporary cooling. Moreover, by plotting the data points during $t_3$ and $t_1$ in Fig. \ref{rlxzfc}(b), we find a continuity between two regions. During the field-change cycle, we find that the relaxation during $t_2$ is fast with opposite sign but as we return back to $T_0$ with a field $h=0.01$ again applied, the magnetization again comes back to the level as before temporary cooling and field change and the relaxation curves during $t_1$ and $t_3$ are continuous (Fig. \ref{rlxzfc}(c)). These results validate the memory effects and also show that the relaxation at lower temperatures has no influence on the states at higher temperatures. Finally during positive-temperature cycle, raising the temperature enhances the number of dynamically active particles which are taking part in relaxation process. The particles which were frozen at $T_0$ become dynamically active at increased temperature and polarize themselves during $t_2$ interval. As we return back to initial temperature $T_0$, these particles remain frozen in polarized state. Hence the relaxation curves during $t_1$ and $t_3$ show an abrupt jump in magnetization as shown in Fig. \ref{rlxzfc}(d).

\begin{figure*}
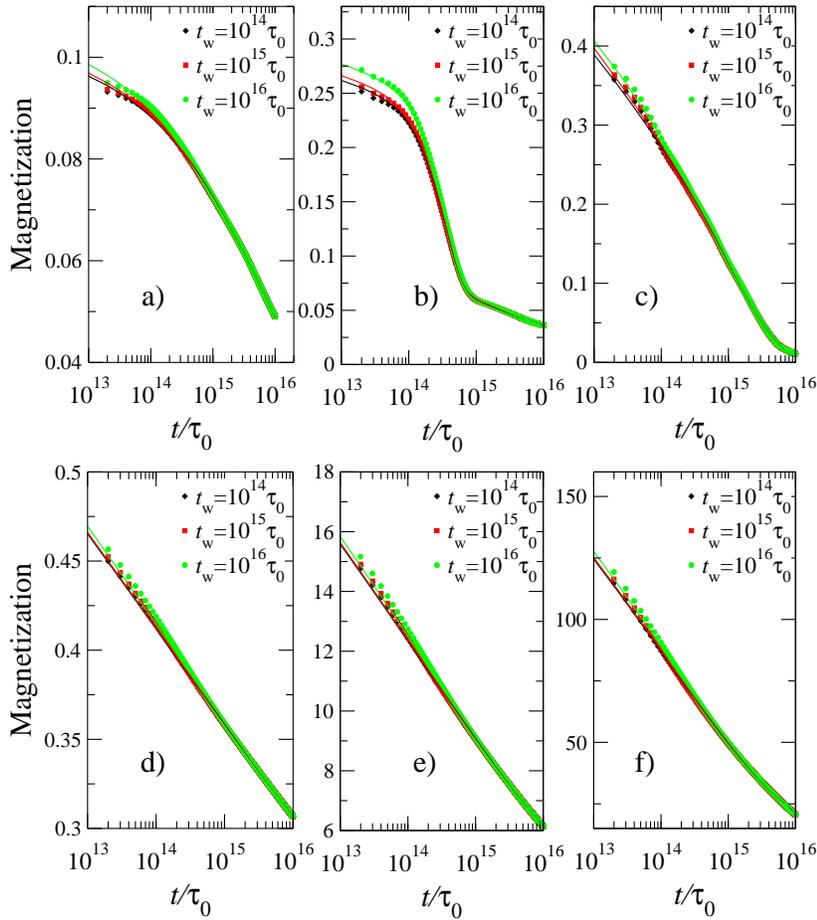

\hspace*{-1.2cm}
\centering
\begin{tabular}{cc}
\epsfig{file=af_aging_fit.eps,width=0.6\linewidth,clip=} \\
\epsfig{file=f_aging_fit.eps,width=0.6\linewidth,clip=} 
\end{tabular}
\caption{ Aging effect for waiting times $10^{14} \tau_0$, $ 10^{15} \tau_0$ and $ 10^{16} \tau_0$.  (a), (b) and (c) correspond to antiferromagnetic nanoparticles for \textit {dist.A} ($1.3a_0-5.3a_0$), \textit {dist.B} ($5.3a_0-9.3a_0$) and \textit {dist.C} ($9.3a_0-13.3a_0$). (d), (e) and (f) correspond to ferromagnetic case for respective size distributions. The solid curves in ferromagnetic cases are fits to Eq. \ref{trm} and those in antiferromagnetic cases are fits to Eq. \ref{trm2}. }
\label{aging}
\end{figure*}
We have also performed the relaxation studies in TRM protocol. Here, as shown in Fig. \ref{rlxzfc}(e), the sample is cooled down to $T_0=0.036$ in a magnetic field $h=0.01$. The field is cut off for a zero wait time and the magnetization relaxation is calculated as a function of time. After a time interval $t_1= 10^{14} \tau_0$, again we employ following three different processes to study the relaxation for $t_2=10^{14} \tau_0$.
\begin{itemize}
\item[A]- Negative-temperature cycle: the sample is quenched in zero field to a lower temperature, $T_{\rm{low}}=0.024 $, and the magnetization is considered for interval $t_2=t_1= 10^{14} \tau_0$.
\item[B]- Field-change cycle: the sample is quenched to a lower temperature, $T_{\rm{low}}=0.024$ with the magnetic field $h=0.01$ applied, and the magnetization is considered again for a time $t_2=t_1= 10^{14} \tau_0$.
\item[C]- Positive-temperature cycle: the sample is quenched in zero field to a higher temperature, $T_{\rm{high}}=0.048$, and the magnetization is considered  for a time $t_2=t_1=10^{14} \tau_0$.
\end{itemize}
At last we quench all the above three processes separately to $T_0=0.036$ with zero field to observe the relaxation for interval $2\times10^{14}-3\times 10^{14} \tau_0$.
We have also shown the magnetization during $t_1$ and $t_3$ for the three different processes occurred during $t_2$ in  Fig. \ref{rlxzfc}(f), \ref{rlxzfc}(g) and \ref{rlxzfc}(h). We find that the logarithmic relaxations in Fig. \ref{rlxzfc}(f) and Fig. \ref{rlxzfc}(g) are continuous which again indicates the memory effect. On the other hand during the positive-temperature cycle, as shown in Fig. \ref{rlxzfc}(h), the magnetization is not continuous. This is because the bigger size particles, which were frozen at $T_0$, become dynamically active at higher temperature $T_{\rm{high}}$. These particles depolarize themselves during $t_2$ interval and remain frozen in depolarized state, when temperature is brought back to lower temperature $T_{0}$. 

The above relaxation studies in ZFC and TRM protocols reveal that the negative-temperature cycle and field-change cycle show memory effect, while 
positive-temperature cycle does not imprint memory during the cycle. Thus the memory effect is only due to the fast relaxation of smaller particles which respond to temporary cooling and field change. %Thus magnetization during $t_1$ 
%%%%%%%%%%%%%%%%%%%%%%%%%%%%%%%%%%%%%%%%%%%%%%%%%%%%%%%%%%%%%%%%%%%%
\section{Aging effect}
\label{secaging} 
%%%%%%%%%%%%%%%%%%%%%%%%%%%%%%%%%%%%%%%%%%%%%%%%%%%%%%%%%%%%%%%%%%%%
Aging effect is a well studied phenomenon in spin glass system.\cite{lundgren,ocio1} Recently it has gained a lot of attention in the system of nanoparticles, where slow dynamics study becomes important to characterize both the superspin glass behavior and the superparamagnetism.\cite{sasaki,sahoo1,sahoo2,sahoo3,tsoi} Most of the experiments are performed by measurement of time-dependent ZFC and TRM magnetizations. In a TRM protocol the system is cooled in a field to a base temperature $T_{\rm{base}}$ below blocking temperature $T_{\rm B}$. After a waiting time $t_{\rm{w}}$, the magnetic field is switched off and one observes the relaxation in magnetization. It is found that the time dependence of the magnetization depends on the waiting time $t_{\rm{w}}$. %The most important feature of the time dependent magnetization is an inflection point in the magnetization vs $\rm{log}_{10}(t)$ curve and a corresponding extremum in the relaxation rate $S(t)=\partial M/\partial ln(t)$ for observation time $t\sim t_{\rm{w}}$.\cite{sasaki,sahoo1,sahoo2,sahoo3,tsoi}
 These studies also show that even a noninteracitng system of ferromagnetic nanoparticles can exhibit the aging effect though weak.

\begin{figure*}
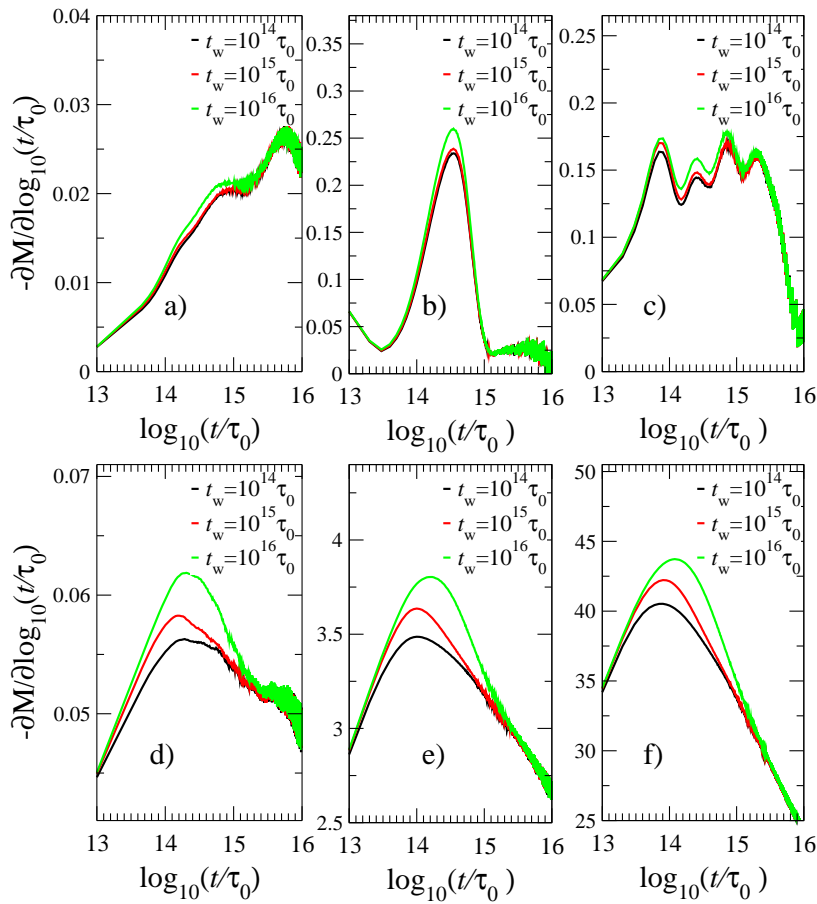

\hspace*{-1.2cm}
\centering
\begin{tabular}{cc}
\epsfig{file=af_rlx.eps,width=0.6\linewidth,clip=} \\
\epsfig{file=f_rlx.eps,width=0.6\linewidth,clip=} 
\end{tabular}
\caption{ Relaxation rates for waiting times $10^{14} \tau_0$, $ 10^{15} \tau_0$ and $ 10^{16} \tau_0$.  (a), (b) and (c) correspond to antiferromagnetic nanoparticles for {\it dist.A} ($1.3a_0-5.3a_0$), {\it dist.B} ($5.3a_0-9.3a_0$) and {\it dist.C} ($9.3a_0-13.3a_0$). (d), (e) and (f) correspond to ferromagnetic case for respective size distributions. A peak can be seen in the ferromagnetic cases, which depends on wait time $t_{\rm{w}}$ and corresponds to $\tau_m$ in Eq. \ref{trm}. Multiple peaks can be seen in antiferromagnetic cases and the relaxation of TRM magnetization is governed by a two step stretched exponential function given by Eq. \ref{trm2}.} 
\label{rlx}
\end{figure*}
 We have studied the aging effect using a polydisperse system of NiO nanoparticles in TRM protocol. Our investigation
is carried out by cooling the system in the presence of magnetic field $h=0.01$ upto base temperature $T_{\rm{base}}=0.024$ and cutting the field off after a wait time $t_{\rm{w}}$ to let the system relax. For all the three size distributions, system of nanoparticles, either ferromagnetic or antiferromagnetic, show wait time dependence. In Figs \ref{aging}(a), (b) and (c), we plot the magnetization of antiferromagnetic nanoparticles with logarithmic time scale for different wait times $10^{14} \tau_0$, $ 10^{15} \tau_0$ and $ 10^{16} \tau_0$ for distributions \textit {dist.A}, \textit{dist.B} and \textit{dist.C} %{\it dist.A}, {\it dist.B} and {\it dist.C}
 respectively. We also plot  relaxation curve for the case of ferromagnetic nanoparticles using same size distributions \textit {dist.A}, \textit{dist.B} and \textit{dist.C} in Figs \ref{aging}(d), (e) and(f). We see that relaxation of antiferromagnetic nanoparticles is qualitatively different than that of ferromagnetic particles.
The time dependence of thermoremenant magnetization in ferromagnetic nanoparticles can be described by a stretched
exponential function
\begin{eqnarray}
\mu(t)=\mu(0) \rm{exp}(-(t/\tau_m)^n),
\label{trm}
\end{eqnarray}
where $\tau_m$ is response time which has a correspondence with the peak position of relaxation rate $S(t)=\partial M/\partial\rm{log}_{10}(t)$ versus $\rm{log}_{10}(t)$ curve.

Best fit of Eq. \ref{trm} to aging data of distribution \textit {dist.A} shows $\tau_m=1.078\times10^{14}\tau_0$ and $n=0.0566$ for waiting time $10^{14} \tau_0$; $\tau_m=1.105\times10^{14}\tau_0$ and $n=0.0569$ for waiting time $10^{15}\tau_0$; and $\tau_m=1.2053\times10^{14}\tau_0$ and $n=0.0582$ for waiting time $10^{16}\tau_0$.
Using same equation to fit the aging data of distribution \textit {dist.B} reveal $\tau_m=1.001\times10^{14}\tau_0$ and $n=0.115$ for waiting time $10^{14} \tau_0$; $\tau_m=1.025\times10^{14}\tau_0$ and $n=0.116$ for waiting time $10^{15}\tau_0$; and $\tau_m=1.12\times10^{14}\tau_0$ and $n=0.119$ for waiting time $10^{16}\tau_0$.
For distribution \textit {dist.C} consisting of bigger sizes, the fitting parameters are $\tau_m=1.0\times10^{14}\tau_0$ and $n=0.192$ for waiting time $10^{14} \tau_0$; $\tau_m=1.028\times10^{14}\tau_0$ and $n=0.194$ for waiting time $10^{15}\tau_0$; and $\tau_m=1.128\times10^{14}\tau_0$ and $n=0.199$ for waiting time $10^{16}\tau_0$.

For all the distributions discussed above, we find a small increase in parameter $n$ with waiting time $t_{\rm{w}}$. Thus parameter $n$ can be useful to quantitatively describe aging effect. %Thus even a superparamagnet can show a small increase in waiting parameter with wait time.
  
However, Eq. \ref{trm} does not satisfy the relaxation of antiferromagnetic nanoparticles. As we can see multiple peaks in relaxation curves shown in Figs. \ref{rlx}(a), (b) and (c), one step exponential decay can not be sufficient to describe the behavior in this case.  We define a two step stretched exponential decay for magnetization of the form
\begin{eqnarray} 
\mu(t)=\mu(0) \left[ \rm{exp}(-(t/\tau_{m_1})^{n_1})+\rm{exp}(-(t/\tau_{m_2})^{n_2}) \right].
\label{trm2}
\end{eqnarray}
The best fit of Eq. \ref{trm2} to magnetization vs $\rm{log}_{10}(t)$ data for various size distributions are listed below.
For  \textit {dist.A}, we find the parameters $\tau_{m_1}=1.03\times10^{13}\tau_0$, $n_1=0.0124$, $\tau_{m_2}=7.282\times10^{15}\tau_0$ and $n_2=0.392$, for waiting time $10^{14} \tau_0$;  $\tau_{m_1}=1.03\times10^{13}\tau_0$, $n_1=0.0157$, $\tau_{m_2}=7.40\times10^{15}\tau_0$ and $n_2=0.385$, for waiting time $10^{15}\tau_0$; and  $\tau_{m_1}=1.05\times10^{13}\tau_0$, $n_1=0.0158$, $\tau_{m_2}=6.52\times10^{15}\tau_0$ and $n_2=0.373$, for waiting time $10^{16}\tau_0$.
A slight increase in $n_1$ and a decrease in $n_2$ with wait time can be seen for this size distribution.
 
For  \textit {dist.B},  fitting parameters are $\tau_{m_1}=3.48\times10^{14}\tau_0$, $n_1=1.603$, $\tau_{m_2}=1.12\times10^{15}\tau_0$ and $n_2=0.183$, for waiting time $10^{14}\tau_0$; $\tau_{m_1}=3.46\times10^{14}\tau_0$, $n_1=1.604$, $\tau_{m_2}=1.0\times10^{15}\tau_0$ and $n_2=0.181$, for waiting time $10^{15}  \tau_0$;  and $\tau_{m_1}=3.53\times10^{14}\tau_0$, $n_1=1.749$, $\tau_{m_2}=1.0\times10^{15}\tau_0$ and $n_2=0.193$, for waiting time $10^{16}\tau_0$.
We see that for \textit{dist.B}, waiting parameters $n_1$ and $n_2$ show increasing trend with waiting time.

However for  bigger size distribution \textit {dist.C}, fitting parameters are $\tau_{m_1}=1.01\times10^{13}\tau_0$, $n_1=0.34$, $\tau_{m_2}=1.25\times10^{15}\tau_0$ and $n_2=0.619$, for waiting time $10^{14}\tau_0$; $\tau_{m_1}=1.02\times10^{13}\tau_0$, $n_1=0.35$, $\tau_{m_2}=1.22\times10^{15}\tau_0$ and $n_2=0.613$, for waiting time $10^{15}  \tau_0$;  and $\tau_{m_1}=1.3\times10^{13}\tau_0$, $n_1=0.329$, $\tau_{m_2}=1.19\times10^{15}\tau_0$ and $n_2=0.606$, for waiting time $10^{16}\tau_0$.
We find that parameters $n_1$ and $n_2$ are decreasing with increasing wait time.

This may easily be understood from a weak dependence of blocking volume on logarithmic observation time. For the case of ferromagnetic nanoparticle, demagnetization of particles with logarithmic time is linear but the same is not true for antiferromagnetic case. We also find that magnetization decays faster for bigger size distribution \textit{dist.C} than smaller sizes in both case ferromagnetic as well as antiferromagnetic nanoparticles.

\section{Conclusions}
\label{secconclusion}
We have studied the effect of size-dependent magnetization fluctuations on the dynamics of the polydisperse system of AFNs by solving two state model analytically. A collection of a few antiferromagnetic nanoparticles has been  has been Numerical calculation of ZFC magnetization shows ripples in the curve which is absent in ferromagnetic particles of same size. These ripples are signature  of size-dependent fluctuations in magnetization and they become more pronounced as heating rate is increased.
The distribution of sizes also play an important role in the time dependent properties of the polydisperse system of nanoparticles. Ripples in ZFC magnetization curve are more highlighted for smaller size distribution and disappear for larger sizes.
 A broad distribution of particle relaxation times arising from the polydispersity is found to be 
responsible for the dynamics of the system of nanoparticles viz superparamagnetic.  
%In order to highlight the behavior of size-dependent magnetization fluctuations, we have also extended the study to two-peak polydispersity.
%The effect of size dependent fluctuations is visible at very low temperatures. 
The memory effect and a weak aging effect has also been observed in a noninteracting polydisperse assembly of nanoparticles for various size distributions. For very small nanoparticles, Memory effect is more in antiferromagnetic case than ferromagnetic case. The situation reverses for bigger size nanoparticles. We have also discussed various relaxation measurements with sudden cooling, heating and removal of fields to validate the memory effects. We have found that the striking memory effects in system of AFNs are indeed originated from polydispersity.
A fitting to aging data in antiferromagnetic nanoparticles shows a two step stretched exponential decay, as contrast to ferromagnetic case, where magnetization show a stretched exponential decay. This can also be confirmed by observed multiple peaks in the relaxation rate versus $\rm{log}_{10}(t)$ curve. In the case of ferromagnetic nanoparticles, aging parameter $n$ shows an increasing trend with waiting time. For antiferromagnetic nanoparticles, depending upon the distribution of sizes, aging parameters $n_1$ and $n_2$ can increase or decrease with waiting time.  
% and wiggles in various relaxation phenomena are the effect of size-dependent magnetization fluctuations.
%  are exhibited in various relaxation phenomena as a wiggle in the curve.
%  \section*{Acknowledgments} 
\section*{Acknowledgments} 
It is pleasure to acknowledge Prof. V. Subrahmanyam for extensive discussions and support during preparation of manuscript. The author also wishes thanks to Prof. K. P. Rajeev for useful discussions, Sudhakar Pandey and Naveen Kumar Singh for help.
 The financial support provided by the Council of 
Scientific and Industrial Research CSIR, Government of India is highly appreciated. 
%\section*{References}

%\begin{figure}[h]
%\vspace*{1.2cm}
%\begin{center}
%\includegraphics [angle=0,width=10cm] {r9_13.eps}
%\caption{ a) r9-13}
%\end{center}
%\label{fig1}
%\end{figure}

%\begin{figure}[h]
%\begin{center}
%\includegraphics [angle=0,width=8cm] {r5_9cmp.eps}
%\caption{ aging of antiferromagnetic NP for differnt temperatures}
%\end{center}
%\label{distcmp1}
%\end{figure}


\section*{References}
\begin{thebibliography}{48}
%Carlip S and Vera R 1998 {\it Phys. Rev.} D {\bf 58} 011345 
\bibitem{weller} D.~Weller and A.~Moser, IEEE Trans. Magn. {\bf 35}, 4423 (1999).
\bibitem{richter} H.~J.~Richter J. Phys. D {\bf 40}, R149 (2007).
\bibitem{berry} C.~C.~Berry and A.~S.~G.~Curtis,  J. Phys. D {\bf 36}, R198 (2003).
%\bibitem{fiorani} {\it Surface Effects in Magnetic Nanoparticles}, edited by D.~Fiorani (Springer, New 
%York, 2005)
\bibitem{fiorani} Surface Effects in Magnetic Nanoparticles, edited by D.~Fiorani (Springer, New 
York, 2005).
  \bibitem{kodama} R.~H.~Kodama, S.~A.~Makhlouf, and A.~E.~Berkowitz, Phys. Rev.
  Lett. {\bf 79}, 1393 (1997). 
\bibitem{kodama1} R.~H.~Kodama, A.~E.~Berkowitz, Phys. Rev. B {\bf 59}, 6321 (1999)  

%\bibitem{kodama1} Kodama R H and Berkowitz A E, 1999 {\it Phys. Rev.} B {\bf 59} 6321
  \bibitem{jonsson1} T.~Jonsson, J.~Mattsson, C.~Djurberg, F.~A.~Khan, P.~Nordblad, and P.~Svedlindh Phys. Rev.
Lett. {\bf 75}, 4138 (1995).
\bibitem{jonsson2} P.~J\"{o}nsson, M.~F.~Hansen, and P.~Nordblad, Phys. Rev. B {\bf 61}, 1261 (2000).
\bibitem{sahoo1} S.~Sahoo, O.~Petracic, Ch.~Binek, W.~Kleemann, J.~B.~Sousa, S.~Cardoso, and P.~P.~Freitas, Phys. Rev. B {\bf 65}, 134406 (2002). 

\bibitem{dormann} J.~L.~Dormann, R.~Cherkaoui, L.~Spinu, M.~Nogues, F.~Lucari, F.~D'Orazio, A.~Garcia, E.~Tronc, and J.~P.~Jolivet, J. Magn. Magn. Mater. {\bf 187}, L139 (1998).
\bibitem{labarata} X.~Batlle and A.~Labarta, J. Phys. D {\bf 35}, R15 (2002).
\bibitem{jonsson3} P.~J\"{o}nsson, Adv. Chem. Phys. {\bf 128}, 191 (2004).
\bibitem{djuberg} C.~Djurberg, P.~Svedlindh, P.~Nordblad, M.~F.~Hansen, F.~B\o{}dker, and S.~M\o{}rup, Phys. Rev. Lett. {\bf 79} 5154 (1997). 
\bibitem{malay2} M.~Bandyopadhyay and J.~Bhattacharya, J. Phys.: Condens. Matter {\bf 18}, 11309 (2006).
\bibitem{garcia1} J.~L.~Garcia-Palacios, Adv. Chem. Phys. {\bf 112}, 1 (2007).
\bibitem{jonsson} T.~Jonsson, P.~Svedlindh, and M.~F.~Hansen, Phys. Rev. Lett. {\bf 81}, 3976 (1998). 
\bibitem{mamiya} H.~Mamiya, I.~Nakatani and T.~Furubayashi, Phys. Rev. Lett. {\bf 82}, 4332 (1999). 
\bibitem{sahoo2} S.~Sahoo, O.~Petracic, W.~Kleemann, P.~Nordblad, S.~Cardoso, and P.P.~Freitas, Phys. Rev. B {\bf 67}, 214422 (2003). 
\bibitem{sun} Y.~Sun, M.~B.~Salamon, K.~Garnier, and R.~S.~Averback, Phys. Rev. Lett. {\bf 91}, 167206 (2003). 
\bibitem{sahoo3} O.~Petracic, X.~Chen, S.~Bedanta, W.~Kleemann, S.~Sahoo, S.~Cardoso, and P.~P.~Freitas, J. Magn. Magn. Mater. {\bf 300}, 192 (2006).
\bibitem{zheng} R.~K.~Zheng, H.~Gu, and X.~X.~Zhang, Phys. Rev. Lett. {\bf 93}, 139702 (2004).
\bibitem{sasaki} M.~Sasaki, P.~E.~J\"{o}nsson, H.~Takayama and H.~Mamiya, Phys. Rev. B {\bf 71}, 104405 (2005).
\bibitem{tsoi} G.~M.~Tsoi, L.~E.~Wenger, U.~Senaratne, R.~J.~Tackett, E.~C.~Buc, R.~Naik, P.~P.~Vaishnava, and V.~Naik, Phys. Rev. B {\bf 72} 014445 (2005).
\bibitem{malay} M.~Bandyopadhyay and S.~Dattagupta, Phys. Rev. B {\bf 74}, 214410 (2006).
\bibitem{wang} W.~J.~Wang, J.~J.~Deng, J.~Lu, B.~Q.~Sun, and J.~H.~Zhao, Appl. Phys. Lett. {\bf 91}, 202503 (2005), W.~J.~Wang, J.~J.~Deng, J.~Lu, B.~Q.~Sun, X.~G.~Wu, and J.~H.~Zhao, J. Appl. Phys. {\bf 105}, 053912 (2009).  
\bibitem{du} J.~Du, B.~Zhang, R.~K.~Zheng and X.~X.~Zhang, Phys. Rev. B {\bf 75}, 014415 (2007).
\bibitem{suzuki} M.~Suzuki, S.~I.~Fullem, I.~S.~Suzuki, L.~Wang and Chuan-Jian Zhong,  Phys. Rev. B {\bf 79} 024418 (2009). 
\bibitem{winkler} E.~Winkler, R.~D.~Zysler, M.~Vasquez Mansilla, and D.~Fiorani,  Phys. Rev. B {\bf 72}, 132409 (2005). 
\bibitem{makhlouf} S.~A.~Makhlouf, F.~T.~Parker, F.~E.~Spada and A.~E.~Berkowitz, J. Appl. Phys. {\bf 81}, 5561 (1997). 
 \bibitem{tiwari} S.~D.~Tiwari and K.~P.~Rajeev, Phys. Rev. B {\bf 72}, 104433 (2005). 
 \bibitem{sunil} S.~K.~Mishra, V.~Subrahmanyam, {\it arXiv:0806.1262v3.} (2008).
\bibitem{brown} L.~N\'eel, Ann. Geophys. C.N.R.S. {\bf 5}, 99 (1949); W.~F.~Brown Jr., Phys. Rev. {\bf 130}, 1677 (1963).
 \bibitem{andersson} J.~-O~Andersson, C.~Djurberg, T.~Jonsson, P.~Svedlindh and P.~Nordblad, Phys. Rev. B {\bf
 56}, 13983 (1997). 
\bibitem{ulrich} M.~Ulrich, J.~Garcia-Otero, J.~Rivas, and A.~Bunde Phys. Rev. B {\bf 67}, 024416 (2003).
\bibitem{garcia} J.~Garcia-Otero, M.~Porto, J.~Rivas, and A.~Bunde, Phys. Rev. Lett. {\bf 84}, 167 (2000). 
\bibitem{neel} L.~Neel in {\it Low Temperature Physics}, edited by C.~DeWitt, B.~Dreyfus and 
P.~G.~DeGennes (Gordon and Beach, London, 1962), p.~411.
%\bibitem{neel} N\'eel L in {\it Low Temperature Physics}, edited by C.~DeWitt, B.~Dreyfus and, 
%P.~G.~DeGennes (Gordon and Beach, London, 1962) p~411
\bibitem{jacob} I.~S.~Jacobs and C.~P.~Bean, in {\it Magnetism}, edited by G.~T.~Rado and 
H.~Suhl (Academic Press, New York, 1963), Vol. III, p.~294.
%\bibitem{jacob} I.~S.~Jacobs and C.~P.~Bean, in {\it Magnetism}, edited by G.Rado G T and 
%Suhl H (Academic Press, New York, 1963) vol. III p~294
\bibitem{richard2}  J.~T.~Richardson, D.~I.~Yiagas, B.~Turk, K.~Forster, and M.~V.~Twigg, J. Appl. 
Phys. {\bf 70}, 6977 (1991). 

\bibitem{schuele} W.~J.~Schuele and V.~D.~Deetscreek, J. Appl. Phys. {\bf 33}, 1136 (1962). 

\bibitem{trohidou} K.~N.~Trohidou, X.~Zianni and A.~J.~Blackman, IEEE Trans. Magn. {\bf 34}, 1120 (1998). 
 \bibitem{zianni}  X.~Zianni and K.~N.~Trohidou, J. Appl. Phys. {\bf 85}, 1050 (1999). 

 \bibitem{zysler} R.~D.~Zysler, E.~Winkler, M.~Vasquez Mansilla and D.~Fiorani   Physica B {\bf384} 277 (2006).  
 
\bibitem{buhrman} C.~G.~Granqvist and R.~A.~Buhrman, J. Appl. Phys. {\bf 47}, 2200 (1976).
\bibitem{hutchings} M.~T.~Hutchings and E.~J.~Samuelsen, Phys. Rev. B {\bf 6}, 3447 (1972).
\bibitem{bean} C.~P.~Bean and J.~D.~Livingston, J. Appl. Phys. {\bf 30}, S120 (1959).
\bibitem{binder} K.~Binder and D.~W.~Heermann D W, {\it Monte Carlo Simulations in Statistical Physics}, Springer Series in Solid State Science Vol. 80 (Springer, Berlin, 1992) 2nd ed
%\bibitem{sunil2} Mishra S K, Subrahmanyam V {\it unpublished}
\bibitem{bisht}	V.~Bisht and K.~P.~Rajeev, arXiv:0909.1391v2 [cond-mat.mes-hall]
\bibitem{malay1} S.~Chakraverty, M.~Bandyopadhyay, S.~Chatterjee, S.~Dattagupta, A.~Frydman, S.~Sengupta, and P.~A.~Sreeram, Phys. Rev. B {\bf 71}, 054401 (2005).
\bibitem{nowak1} U.~Nowak, R.~W.~Chantrell, and E.~C.~Kennedy, Phys. Rev. Lett. {\bf 84}, 163 (2000), D.~Hinzke and U.~Nowak, Phys. Rev. B {\bf 61}, 6734 (2000).
\bibitem{lundgren}  L.~Lundgren, P.~Svedlindh, P.~Nordblad and O.~Beckman, Phys. Rev. Lett. {\bf 51}, 911 (1983).
\bibitem{ocio1} M.~Ocio, M.~Alba and J.~Hammann, J. Phys. (France) Lett. {\bf 46}, L1101 (1985); M.~Ocio, M.~Alba and J.~Hammann, Europhys. Lett. {\bf 2}, 45 (1986).
\end{thebibliography}
\end{document}